\documentstyle[12pt,epsfig]{article}

\advance\textheight by \topskip
\newcommand{\be}{\begin{equation}}
\newcommand{\ee}{\end{equation}}
\newcommand{\br}{\begin{eqnarray}}
\newcommand{\bea}{\begin{eqnarray}}
\newcommand{\beanon}{\begin{eqnarray*}}
\newcommand{\er}{\end{eqnarray}}
\newcommand{\eea}{\end{eqnarray}}
\newcommand{\eeanon}{\end{eqnarray*}}
\newcommand{\ba}{\begin{array}}
\newcommand{\ea}{\end{array}}
\newcommand{\bi}{\begin{itemize}}
\newcommand{\ei}{\end{itemize}}
\newcommand{\bn}{\begin{enumerate}}
\newcommand{\en}{\end{enumerate}}
\newcommand{\bc}{\begin{center}}
\newcommand{\ec}{\end{center}}

\newcommand{\ar}{\rightarrow}

\newcommand{\Dir}{\kern -6.4pt\Big{/}}
\newcommand{\Dirin}{\kern -10.4pt\Big{/}\kern 4.4pt}
\newcommand{\DDir}{\kern -7.6pt\Big{/}}
\newcommand{\DGir}{\kern -6.0pt\Big{/}}
\def\Ord{\buildrel{\scriptscriptstyle <}\over{\scriptscriptstyle\sim}}
\def\OOrd{\buildrel{\scriptscriptstyle >}\over{\scriptscriptstyle\sim}}
\def\sm{\ifmmode{{\cal {SM}}}\else{${\cal {SM}}$}\fi}
\def\mt{\ifmmode{{m_{t}}}\else{${m_{t}}$}\fi}
\def\MH{\ifmmode{{M_{H}}}\else{${M_{H}}$}\fi}
\def\MWpm{\ifmmode{{M_{W^\pm}}}\else{${M_{W^\pm}}$}\fi}
\def\Wpm{\ifmmode{{{W^\pm}}}\else{${{W^\pm}}$}\fi}
\def\pl #1 #2 #3 {{\it Phys.~Lett.} {\bf#1} (#2) #3}
\def\np #1 #2 #3 {{\it Nucl.~Phys.} {\bf#1} (#2) #3}
\def\zp #1 #2 #3 {{\it Z.~Phys.} {\bf#1} (#2) #3}
\def\pr #1 #2 #3 {{\it Phys.~Rev.} {\bf#1} (#2) #3}
\def\prep #1 #2 #3 {{\it Phys.~Rep.} {\bf#1} (#2) #3}
\def\prl #1 #2 #3 {{\it Phys.~Rev.~Lett.} {\bf#1} (#2) #3}
\def\mpl #1 #2 #3 {{\it Mod.~Phys.~Lett.} {\bf#1} (#2) #3}
\def\rmp #1 #2 #3 {{\it Rev. Mod. Phys.} {\bf#1} (#2) #3}
\def\xx #1 #2 #3 {{\bf#1}, (#2) #3}

\begin{document}
\tolerance=100000
\thispagestyle{empty}
\setcounter{page}{0}

\begin{flushright}
{\large Cavendish--HEP--97/04}\\
{\rm June 1997\hspace*{.5 truecm}}\\ 
\end{flushright}

\vspace*{\fill}

\begin{center}
{\large \bf Single-top production at future $ep$ colliders}\\[2.cm]
{\large Stefano Moretti$^1$ and Kosuke Odagiri\footnote{E-mail: 
moretti, odagiri@hep.phy.cam.ac.uk}}\\[.3 cm]
{\it Cavendish Laboratory, University of Cambridge,}\\
{\it Madingley Road, Cambridge, CB3 0HE, United Kingdom.}\\[1cm]
\end{center}

\vspace*{\fill}

\begin{abstract}{\normalsize\noindent The production
of top quarks in single mode  
at future $ep$ colliders is studied, the attention being mainly 
focused to the case of the proposed 
LEP$\oplus$LHC collider. We are motivated to reanalyse such a process 
following the discovery of the top quark at Fermilab. Thanks
to the measurement of its mass one is now able to establish more accurately 
the relevance of single top production
 for itself and for many other processes to which it may
act as a background. In addition, the recent improvement
 of our knowledge of the 
quark and gluon dynamics inside the proton now allows one to pin down the
dependence of single top production on the partonic  structure
functions. Both the leptonic and hadronic decay channels of the top quark are 
studied and compared to the yield of the corresponding irreducible background 
in presence of $b$-tagging.}\end{abstract}

\vspace*{\fill}
\newpage

\section*{1. Introduction} 

Now that the Fermilab experiments have clearly assessed the existence of the 
heaviest quark of the Standard Model (SM) \cite{discovery} and given 
a rather accurate measurement of its mass \cite{value}, many of the theoretical
calculations carried out in the previous years need to be updated to the
current value of this fundamental parameter. In this paper, we
turn our attention to the case of top production in single mode at future 
electron(positron)-proton colliders. 
As a further motivation for our revision we
put forward the fact that a huge amount of data improving our knowledge of 
the parton distribution functions (PDFs) has been produced in the years following
the early studies of top phenomenology at $ep$ machines (see, e.g., 
Ref.~\cite{LHC}), along with more detailed treatments of the dynamics of
heavy quarks inside the proton. Therefore, the error associated with 
the partonic 
behaviour in the initial state should be at present significantly smaller
than in the past. 
Finally, the 
reduction of the theoretical uncertainty in single top production 
also implies that other effects, such as those due to the irreducible
backgrounds, need now to be incorporated in more 
detailed phenomenological analyses.
We calculate these effects here 
for the first time, in both the hadronic
and leptonic top decay channels.

In order to illustrate the particular relevance of  single top processes 
in electron(positron)-proton annihilations we remind the 
readers of the motivations for higher energy $ep$ experiments.

Firstly, such colliders will be an obvious and unrivaled testing ground
for QCD at very low Bjorken $x$ \cite{LHC_phys}, in exploring
the structures of both the proton and the photon \cite{photon} at the TeV
scale, taking over the presently running HERA machine \cite{HERA}. 
In connection with this point, we will show that the single top quark process 
discussed here can be be useful in understanding the phenomenology of the PDF
of the bottom quark.

Secondly, and particularly in the case of the proposed LEP$\oplus$LHC
collider \cite{LHC}, they will be able to 
search for the Higgs boson $\phi$ of the
Standard Model \cite{LHC_higgs} (or the lightest neutral Higgs boson of
the Minimal Supersymmetric Standard Model, MSSM) in the intermediate mass
range 90 GeV $\Ord M_\phi\Ord$ 130 GeV \cite{higgs1,higgs2}, in the case it
may not be accessible at the LHC nor the existing colliders (see Ref.
\cite{higgs_review} for discussions on this point). As was
discussed in \cite{LHC_higgs}, the jet background due to the top
quark will be large if only a single $b$-tagging is implemented in
identifying the Higgs boson decay in the most favoured channel
$ep\rightarrow\nu_eW^+W^-X\rightarrow\nu_e\phi X\rightarrow\nu_eb\bar{b}X$
\cite{LHC_higgs}.

Thirdly, the r{\^{o}}le of an $ep$ machine will be complementary to those of
$e^+e^-$ (i.e., NLC) and $pp$ (i.e., LHC) colliders in the search for New 
Physics such as
leptoquarks, excited leptons, low mass sleptons, doubly charged Higgs
bosons, and new vector bosons (see \cite{LHC_phys} and the references
therein). Many of these processes have neutral current-type interactions
of the form $eq \ar eq$, and so the single top quark process, when the top
quark decays leptonically to a bottom quark, a positron and an electron
neutrino, is a potentially dangerous background which should be included
in experimental simulations. As for the possibility of exotic top quark
decays, the study of the Supersymmetric two-body decay modes is a straightforward
extension of this project \cite{charged_higgs} and will be carried out elsewhere
\cite{elsewhere}.

Although the physics potential of a higher energy $ep$ machine is
suppressed compared to a $pp$ one by the reduced centre-of-mass (CM) energy
and luminosity, we stress its allure in the suppression of the 
initial state QCD
noise, which allows for a cleaner environment to study the physics of the
TeV scale, possibly before an NLC will be in operation \cite{ee500}. We
also mention that the physics of $ep$ colliders, in conjunction with
the discussed possibility of their running in the $\gamma p$ mode
\cite{gamma_p}, has been recently under renewed and active discussion
\cite{ankara}. 

The production of top quarks at future $ep$ colliders \cite{LHC} has been
studied in the context of top quark searches at LEP1$\oplus$LHC
\cite{LHC_phys}, during the 1990 Aachen Workshop. A detailed
study was presented in the corresponding Proceedings \cite{LHC_top}.
There, the two following channels were investigated \cite{wg}:
\begin{equation}\label{CC}
\mbox{CC:}\qquad\qquad W^\pm + g\ar t b,
\end{equation}
\begin{equation}\label{NC}
\mbox{NC:}\qquad\qquad \gamma,Z + g\ar t\bar t,
\end{equation}
via charged (CC) and neutral current (NC) scatterings of
an off-shell gauge boson against a gluon, the former being produced via
bremsstrahlung off the incoming electron(positron) and the latter being
extracted from the proton. In general, the CC channel dominates
over the NC one, due to the larger phase space available.
For $m_t=175$ GeV \cite{value}, the suppression is more than one order
of 
magnitude at the TeV scale \cite{LHC_top}. (Indeed, this is the reason why we
will concentrate on $W^\pm g$ fusion only.)
In Ref. \cite{LHC_top}, also a detailed signal-to-background analysis
was carried out, in both the (semi)leptonic and hadronic top decay
channels. 

The Feynman diagrams describing reaction (\ref{CC}) induced, e.g., 
by positron beams, can be found
in Figure 1a, where the top is considered on-shell.
As the bottom mass is small compared to $m_t$,
the dominant contribution to the total cross section comes from diagram
1 of Figure 1a, when the final $\bar b$ quark is
collinear with the incoming gluon. The collinear divergences
are however regulated by the finite value of the bottom mass and 
manifest themselves by means of contributions of the form
$L=\alpha_s(\mu^2)
\log (\mu^2/m_b^2)$, with $\mu^2\sim{\mathaccent94{s}}$, being 
${\mathaccent94{s}}$
the CM energy at the `partonic' level $e^+ b$ and $\alpha_s$  the strong
coupling constant.

Such logarithms are rather large, thus terms of the form 
$\alpha_s^n(\mu^2)\log^n(\mu^2/m_b^2)/n!$
have to be resummed to all orders in perturbation theory
\cite{resum} in order to compute the cross section reliably. This can be done 
by introducing a $b$ parton distribution, $f_b(x,\mu^2)$, in terms of
the Dokshitzer-Gribov-Lipatov-Altarelli-Parisi (DGLAP) splitting function
\be\label{split}
P_{bg}(z)=\frac{1}{2}[z^2+(1-z^2)]
\ee
of a gluon into $b\bar b$ pairs ($z$ being the fractional energy
carried away by the (anti)quark).
In fact, bottoms are not valence quarks, rather they materialise
once the energy scale $\mu$ of the evolution reaches their production 
`threshold'
at a given value $\mu_b\sim m_b$. The function $P_{bg}(z)$ is indeed the `coefficient 
function' 
of the logarithmically enhanced term. The $b$ structure function then 
evolves with $\mu$ according to the DGLAP equations, 
from an  initial condition of the sort, e.g., $f_b(x,\mu^2)=0$ if 
$\mu^2\le\mu_b^2$. 

It follows then that single top production and decay via process ({\ref{CC}})
can be conveniently studied by computing the transition amplitude squared
for the reaction (e.g., assuming incoming positron beams) 
\be\label{t}
e^+ b\ar \bar\nu_e t\ar \bar\nu_e bW^{+}\ar \bar\nu_e b f\bar f',
\ee
where $f$ represents a lepton/neutrino or a $u,d,s$ and $c$ quark (produced
in the top decay), 
appropriately convoluted with a $b$ distribution function evaluated
at the adopted scale $\mu^2$. We exploit here this approach. 

In our opinion, such a procedure (in which the parton is $b$) is  
more appropriate than the one exploited in Ref.~\cite{LHC_top} 
(in which the parton is $g$), especially 
at high energies. In fact, we have explicitly verified that for the values
of $\sqrt s_{ep}$ considered here the
dominant contribution to process (\ref{CC})  comes from configurations 
in which the $\bar b$ is collinear
with the incoming gluon. Since it is exactly such emission that is summed
to all orders in perturbation theory in leading and next-to-leading logarithmic
accuracy inside the $b$ structure function when 
$\alpha_s(\mu^2)\log(\mu^2/m_b^2)\sim {\cal O}(1)$ (and this
is clearly the case in our context, e.g., 
when $\mu^2={\hat s}\OOrd m_t^2$), 
our approach will give a more accurate answer.
However, for comparison, we will also show in the present paper 
several rates as produced by the process 
induced by $g\ar b\bar b$ splitting\footnote{Note that 
the complete next-to-leading (NLO) corrections to $W^\pm g/b$ fusion
involving the collinear logarithms as well as the large angle emission and 
the loop diagrams have been recently presented in the $\overline{\mbox{MS}}$
renormalisation scheme \cite{wgnloms}. Such results contradict
earlier ones based on the DIS factorisation scheme \cite{wgnlodis}.
For the case of $ep$ collisions at HERA they amount to approximately 2\%
of the result obtained by means of the $b$ structure function approach,
and they are rather insensitive to collider energies in the TeV range. 
Therefore we expect them to be well under control also at the proposed
LEP$\oplus$LHC, so for the time being 
we do not include them in our calculation. Another reason
for doing so is that we will also be  concerned with the interplay
between the single top signal and the non-resonant irreducible
background, which is here computed at lowest order.}.

In this paper we 
study the single top quark production via $e^+ b$ fusion at various
energies, together with all tree-level irreducible background processes as
shown in Figures 1b, 1c and 1d. Figure 1b corresponds to the case of the
leptonic decays of the $W^\pm$ boson in the signal process,

\be\label{l}
e^+ b\ar \bar\nu_e b \ell^+ \nu_\ell,
\ee
where $\ell=e,\mu,\tau$, whereas   Figure
1c refers to hadronic decays of the $W^\pm$ boson,
\be\label{h_ew} 
e^+ b\ar \bar\nu_e b \ell^+ jj', 
\ee 
where $jj'$ represents
a pair of light quark jets $u\bar{d}$ or $c\bar{s}$. To these 
must be
added the case of the gluon mediated background of Figure 1d,

\be\label{h_qcd}
e^+ b\ar \bar\nu_e b \ell^+ jj',
\ee
where $jj'$ again represents a pair of light quark jets.

In addition to these three, if charge measurements of the bottom quarks
prove impractical or impossible, we will have background from processes of
the forms: 

\be\label{b_barl}
e^+ \bar b\ar \bar\nu_e \bar b \ell^+ \nu_\ell,
\ee
\be\label{b_barh_ew}
e^+ \bar b\ar \bar\nu_e \bar b \ell^+ jj'.
\ee

A single $b$-tagging capability is assumed throughout, all results being
linearly proportional to its efficiency. 
The case $\ell=\tau$ assumes that jets coming from
the tau and the quarks will easily be distinguishable. We perform all
calculations for the case of $e^+p$ colliders, although the $e^-p$ case is
precisely identical since the calculations involve no valence quarks and
are therefore invariant under the exchange $e^+\leftrightarrow e^-$.

The plan of this paper is as follows. In Section 2, we describe the
methods we adopted in the calculations of the signal and background
processes. In Section 3, we present and discuss our results. Section 4 is
a brief summary. 

\section*{2. Calculation} 

The tree-level Feynman diagrams that one needs in order to compute
processes (\ref{l}), (\ref{h_ew}) and (\ref{h_qcd}) are given in
Figures~1b, c and d, respectively. For reaction (\ref{l}) we show the
diagrams for the case $\ell=e$, which is the most complicated. When $\ell=\mu$
or $\tau$ only ten out of the twenty-one diagrams in Figure 1b contribute.
For processes (\ref{h_ew})--(\ref{h_qcd}) the number of
diagrams is independent of the flavour $j$\footnote{We will refer to
process (\ref{l}) as the `leptonic' channel, and to processes (\ref{h_ew})
and (\ref{h_qcd}) as the `hadronic' channels. For the latter cases, we
will further distinguish between `electroweak' (EW) and `strong' (QCD)
production, respectively.}. 

The single top quark signals (\ref{t}) are produced by diagrams 11 in
Figure 1b and 4 in Figure 1c, for leptonic and hadronic $W^\pm$ decays,
respectively. 
The remaining diagrams in Figures 1b and 1c represent the `irreducible'
background to single-top production and decay. Reaction (\ref{h_qcd}) does
not contribute to the signal at all, but only to the background.

Graphs in Figures 1b, c and d refer to the case of $e^+b$ fusion, i.e., to
the scattering of a positron and a bottom quark, the latter being
extracted from the incoming proton beam. As mentioned earlier on,
we have treated the bottom quark
as a constituent of the proton with the appropriate momentum fraction
distribution $f_{b}(x,\mu^2)$, as given by our partonic structure
functions. It can be noted that the bottom antiquark is also present
inside the proton with an equal probability.  When calculating rates for
single-top production at $e^+p$ colliders, diagrams initiated by bottom
antiquarks must also be considered. However, as long as the deep
inelastic scattering of the proton takes place against a positron, such
graphs do not produce a resonant top quark. The topologies of these bottom
antiquark initiated graphs are easily deducible from those in
Figures 1b and c.
From the point of view of top quark studies, these act as additional
backgrounds. Their production rates will be different from the case of
$e^+b$ fusion if the CM energy at partonic level (i.e.,
$\sqrt{\mathaccent94{s}}$) spans the top quark production threshold. In
contrast, the cross sections due to $e^+\bar b$ initiated diagrams and
proceeding via QCD interactions are identical to the yields of reaction
(\ref{h_qcd}) and the actual graphs are the same as those in Figure 1d,
apart from the trivial operation of reversing the bottom quark
line\footnote{Note that for the case of $e^-p$ scattering things work in a
complementary way, the resonant top antiquarks being produced by incoming
$\bar b$ partons. Indeed, as bottom (anti)quarks are produced inside the
nucleon via a $g\ar b\bar b$ splitting (that is, they are sea partons), no
differences occur in the deep-inelastic dynamics of the above processes if
antiproton beams are considered. Although we study positron-proton
colliders here, our discussions are transposable to all the other cases.}.

The possibility of the `$2b$' charged current processes $e^+q\rightarrow
\bar\nu_eb\bar bq$ (where $q$ is $d$, $\bar u$, $s$ or $\bar c$) being
mistagged as a single $b$ event and acting as background to the hadronic
channel can not be neglected, even when the $b$-tagging efficiency is
high. If the latter is denoted by $\epsilon_b$,
then the probability of misidentification is given by
$2\epsilon_b(1-\epsilon_b)$, assuming no correlations between the two
$b$-taggings. Thus the suppression of the $2b$ background with respect to
the single $b$ events is  $2(1-\epsilon_b)$. As our
investigations concern mainly the top quark signal process (\ref{t}), 
the complete analysis of single bottom quark processes being outside the
scope of our present study, we content ourselves with an estimate of the
degree to which this additional `irreducible'
background could affect the top quark and $W^\pm$ boson mass
reconstruction (the $2b$ background does not contribute to the leptonic
case). Our explicit calculation, using the methods explained below, shows
that after cuts in the reconstructed top quark and $W^\pm$ boson masses are
introduced, the cross section of the $2b$ process is of the same order
as that of reaction (\ref{h_qcd})
(differing only by 10\% at the LEP$\oplus$LHC energies), thus being
quite small in the end (see Section 3.). 

To calculate the squared amplitudes for processes (\ref{l})--(\ref{h_qcd}) 
we have used the {\tt FORTRAN} packages MadGraph \cite{tim} and HELAS
\cite{HELAS}. The codes produced have been carefully checked for gauge and
BRS \cite{BRS} invariance at the amplitude squared level. The
multi-dimensional integrations over the phase spaces have been performed
numerically using the Monte Carlo routine {VEGAS} \cite{VEGAS}, after
folding the partonic differential cross sections with the appropriate
quark densities. The programs that we have produced have been run on
a DEC 3000 Model 300 alpha-station, on which the evaluation of, e.g.,
10$^6$ events took some 14 minutes of charged CPU time to produce a cross
section at the level of percent accuracy in the case of process (\ref{l})
for the sum of the two contributions $\ell=e$ and $\mu$ (the latter being
equivalent to the case $\ell=\tau$): that is, for the channel involving
the largest number of diagrams and the most complicated resonance
structure. 

All the codes implemented are available from the authors upon request. To
allow for a prompt evaluation of single-top rates at any energy and for
any choice of selection cuts, we have also calculated the amplitude
squared of process (\ref{t}) analytically, including top width effects. In 
the leptonic case, and assuming all lepton and neutrino masses to be zero, 
it reads as follows\footnote{The analytic expression for process (\ref{CC}),
also involving the decay currents,
can be found in Ref.~\cite{keith}. We have checked our ME for
the gluon induced process against that given in Ref.~\cite{keith} in
the appropriate configuration (i.e., for $p\bar p$ collisions) and 
found perfect agreement.}:
\vskip0.75cm\noindent
\[
\overline{|M_{e^+ b\ar \bar\nu_e b' \ell^+ \nu_\ell }|}^2
= 2 (4\pi\alpha_{em}/\sin^2\theta_W)^4|P_{W^*}|^2|P_W|^2|P_t|^2
p_b\cdot p_{\bar\nu_e} (-p_e\cdot p_\ell p_t^2 + 
2 p_e\cdot p_t p_\ell \cdot p_t)
\]
with
\[P_{W^*}=1/(p_{W^*}^2-M_{W^\pm}^2),
P_W=1/(p_W^2-M_{W^\pm}^2+iM_{W^\pm}\Gamma_{W^\pm}),
P_t    =1/(p_t^2-m_t^2+im_t\Gamma_t)\]
\[p_{W^*}=p_e-p_{\bar\nu_e},\qquad
p_W=p_\ell+p_{\nu_\ell},\qquad
p_t=p_b+p_{W^*}.\]
\vskip0.75cm
\noindent
In the hadronic case, again assuming zero light quark masses, the above
formula needs to be multiplied by the colour factor 3, and $\ell^+$ and
$\nu_\ell$ replaced by $\bar{d}$($\bar{s}$) and $u$($c$) respectively. 

As the default set of PDFs we used the
NLO set MRS(A) \cite{MRSA}. However, as one of the
motivations of this study is to investigate the dependence of process
(\ref{t}) on the evolution of the structure
functions of bottom quarks inside the proton, we have produced our results
for other 23 recent NLO PDFs which give
excellent fits to a wide range of deep inelastic scattering data and to
others from different hard scattering processes (see the original
references for details). These are the packages MRS(A', G, J, J', R1, 
R2, R3, R4), MRS(105, 110, 115, 120, 125, 130), MRRS(1,2,3)
and CTEQ(2M, 2MS, 2MF, 2ML, 3M, 4M) \cite{MRSA,MRSG,mrs96fit,MRSalphas,
MRSJ,MRSCHM,CTEQ2,CTEQ3,CTEQ4}. Note that in each case the
appropriate value of 
$\Lambda_{\mbox{\tiny{QCD}}}\equiv
\Lambda^{(n_f)}_{\overline{\mbox{\tiny{MS}}}}$
was used. In particular, for the MRS(A) set we
adopted $\Lambda^{(4)}_{\overline{\mbox{\tiny{MS}}}}=230$~MeV. 
Unless otherwise stated, the QCD strong coupling constant
$\alpha_s$ entering explicitly in the production cross section of process
(\ref{h_qcd}) and implicitly in the PDFs was in general evaluated at
two-loop order
at the scale
$\mu=\sqrt{\mathaccent94{s}}$. The same choice has been made for the scale
of the structure functions.
The spread of the results as obtained from the different packages
with respect to our default MRS(A) value (rather than the errors of the
numerical integrations) can be taken as a possible estimate of the uncertainty of our
predictions throughout the all paper\footnote{We have verified that
differences in the results similar to those obtained in case of process
(\ref{t}) also occur for the complete tree-level reactions
(\ref{l})--(\ref{h_qcd}).}.

The bottom quark sea distributions
are not measured by experiment, but are obtained from the gluon
distributions splitting into $b\bar b$ pairs by using the
DGLAP evolution equations
\cite{DGLAP}. Therefore, the $b$ structure functions are different from the
light quark distributions, which do need to be measured as they involve
non-perturbative QCD, for which a consistent theoretical framework does
not exist. In contrast, the PDFs of $b$ quarks evolve at energies
of the order of the fermion mass $m_b$ or larger, so that their dynamics
can be calculated by using the well assessed instruments of perturbative 
QCD. That is, given the PDFs of the gluon and of the light quarks, those
of the $b$ are precisely determined, as they do not contain any free 
parameters (apart from $m_b$, of course).

We think that by the time that $ep$ colliders at the TeV scale
will begin to be operative, the uncertainty on the gluon distributions
at medium and small $x$ may be  expected to be significantly smaller than
at present, principally due to forthcoming 
improved measurements of the small $x$  
deep inelastic structure functions at HERA, and of large $p_T$ jet and prompt 
photon production at the $p \bar p $ (Di-)Tevatron at Fermilab and the $pp$
LHC at CERN (the latter being scheduled to start running 
around 2005)\footnote{In fact,  the typical $x$ values probed via process 
(\ref{t}), e.g., at the LEP$\oplus$LHC, are  of the order 
$m_t^2/{\hat s}$ or
more, that is above 10$^{-2}$, where the gluon density is already 
well known at present.}. Therefore, detailed studies of single top events 
produced in electron-proton collisions will  
allow one to constrain the
error related to the dynamics of the $g\ar b\bar b$ splitting
in the DGLAP evolution.
In fact, we expect the experimental information on $b$ structure 
functions as collected at the end of the HERA, Fermilab and LHC epoch 
to be rather poor, if not inexistent. On the one hand, at the CM 
energy typical of the $ep$ 
accelerator now running at DESY ($\sqrt s_{ep}=314$ GeV) 
the content of $b$ quark inside the scattered
hadron is very much suppressed per se \cite{EHLQ}. On the other hand, at both 
the 
Tevatron ($\sqrt s_{p\bar p}=1.8-2$ TeV) and LHC ($\sqrt s_{pp}=10-14$ TeV) 
the study of $b$ induced processes inevitably proceeds 
through either the production of
top quarks in single mode \cite{wuki}, whose signatures suffer from a huge
background due to $t\bar t$ production via $q\bar q$ and $gg$ fusion, or
via pure QCD interactions, biased by a large amount of light quark and
gluon jet noise.
These two problems can in principle be solved by future $ep$ colliders.
Firstly, they will be operating at the TeV scale thus allowing for a
very much enhanced content of initial $b$ quarks, which can be probed in
the `kinematically' more defined context of 
a DIS process of an electron(positron) against a proton.
Secondly, as discussed previously, the single top mode via $e^\pm p$ collisions
has a much larger cross section than top-antitop production induced by 
$\gamma g$ and/or $Z g$ fusion \cite{LHC_top}. 

As for the dependence of 
process (\ref{t}) that
one expects on the different $b$ structure functions, 
it is worth reminding the reader some peculiar features of the sets 
considered here. For starting, whereas in the earlier MRS sets (excluding
MRRS(1,2,3)) the bottom density is set to zero below threshold (i.e., 
$f_b(x,\mu^2)=0$ for $\mu^2<\mu_b^2$) and for $\mu^2>\mu_b^2$ the bottom 
distribution is evolved assuming a massless quark, $m_b=0$, the most recent
ones (i.e., MRRS(1,2,3) \cite{MRSCHM}) implement a formulation which allows
heavy quarks mass effects to be explicitly incorporated in both the coefficient
and splitting functions in the parton evolution equations and the two
regions $\mu^2\approx\mu_b^2$ and 
$\mu^2\gg\mu_b^2$ can be treated consistently.
Indeed, an alternative approach 
is to treat the bottom as 
a massless parton above $\mu^2=\mu_b^2$ \cite{Alter}: that is, the mass
effects are neglected in the splitting functions, although they are included
in the coefficient function at NLO. The difference between the two techniques
clearly resides in the fact that the neglected $m_b^2$ effects give
NLO contributions during the DGLAP evolution\footnote{Note that a 
third approach exists in literature \cite{PGF}, which does not treat
bottoms as partons. For example, $b$ quarks are not present 
in the GRV sets of PDFs \cite{GRV94}.}. (The approach of Ref.~\cite{MRSCHM}
has however been criticised in Ref.~\cite{nomass}, where it was made the point
that the presence or absence of heavy quark masses in the DGLAP 
evolution kernels has no effects on the measurable cross section.)
Furthermore, also the choice
of the threshold $\mu_b$ can vary, being in some instances set at 
$\mu_b^2=m_b^2$ (see, e.g., Ref.~\cite{MRSG}) and in other cases at 
$\mu_b^2=4m_b^2$ (see, e.g., Ref.~\cite{CTEQ3}). 

For consistency with the parameter values adopted in the set MRRS(1,2,3) 
we have used as default for the mass of the bottom quark
$m_b=4.3$ GeV \cite{CL}. (Note that the charm mass $m_c$ 
in the three above packages has been set equal to 1.35, 1.50 and 1.2 GeV, 
respectively: this is however a `dummy' value in the production process
(\ref{t}).) For the top mass we have taken (unless otherwise stated)
$m_t=175$ GeV \cite{value}, whereas for the width $\Gamma_t$ we have
used the tree-level expression \cite{widthtopSM}. 
Leptons and $u, d, s$ (and $c$ as well) quarks were 
considered as massless in processes (\ref{l})--(\ref{h_qcd}). For 
simplicity, we set the 
Cabibbo-Kobayashi-Maskawa (CKM) mixing matrix element of the
top-bottom coupling equal to one, the Standard Model prediction at the
90\% confidence level \cite{CL} being $0.9989\leq |V_{tb}|\leq 0.9993$.
For the gauge boson masses and widths we used $M_{Z}=91.19$ GeV, 
$\Gamma_{Z}=2.50$
GeV, $M_{W^\pm}=80.23$ GeV and $\Gamma_{W^\pm}=2.08$ GeV.
The electromagnetic coupling constant and the weak
mixing angle are $\alpha_{em}= 1/128$ and $\sin^2\theta_W=0.2320$, 
respectively. 

The Higgs boson of the Standard Model enters directly in the diagrams of
Figures 1b (graphs 3 and 16) and c (graph 7), when the bottom quark mass
is retained in the fermion-fermion-scalar vertex. As default value for
the scalar mass we used $M_H=150$ GeV, according to the best $\chi^2$ fit
as obtained from the analysis of the LEP and SLC high precision EW data: 
i.e., $M_H=149^{+148}_{-82}$ GeV \cite{theory}. However, since the
constraints on the Higgs mass are rather weak (a lower bound of 66 GeV
from direct searches \cite{lower} and a 95\% confidence level upper limit
of 550 GeV from the data mentioned \cite{rep}) we studied the $M_H$
dependence of the EW contributions in processes (\ref{l})--(\ref{h_ew}) and
(\ref{b_barl})--(\ref{b_barh_ew}),
and found it negligible (note that the Higgs boson is always produced via
non-resonant channels in those reactions). This is also true for the 
$2b$ process, after the implementation of the selection cuts (see below). 

Finally, as total CM energy $\sqrt s_{ep}$ of the colliding
positron-proton beams we have adopted values in the range between 300 GeV
(i.e., around the HERA value) and 2 TeV. However, we focused our attention
mainly to the case of a possible LEP2$\oplus$LHC accelerator, using a 100
GeV positron beam from LEP2 and a 7 TeV proton one from the LHC, yielding
the value $\sqrt s_{ep}\approx1.7$ TeV in the CM frame of the
colliding particles.

\section*{3. Results}

As emphasised in Section 1, 
we generate the single top quark in the final state
by means of the matrix element for $e^+b$ fusion (\ref{t}) convoluted with 
$b$ structure functions rather than producing the initial $b$ quark
via an exact $g\ar b\bar b$ splitting folded with a gluon density.
However, to 
investigate the differences between the two procedures, we show in 
Figure 2 the total cross section of the signal process (\ref{t}) 
plotted against the CM energy of the $ep$ system along
with the yield of reaction (\ref{CC}) (the latter including top decays and
finite width effects on the same footing as the former)\footnote{Note that
in order to obtain a gauge invariant cross section for process (\ref{CC})
in presence of a finite value of $\Gamma_t$ we need to consider a set
of three diagrams. That is, the two with resonant top production (i.e., those
in Figure 1a with the additional 
decay $t\ar bW^+\ar bf\bar f'$) and a third one
in which the $W^+\ar f\bar f'$ current is attached to the off-shell
fermion propagator in one of the graphs of Figure 1a.}. 

Care must be taken when comparing processes (\ref{CC}) and (\ref{t})
with respect to each other. In fact, one should recall that the corresponding
rates are strongly dependent on the (factorisation) scale $\mu$. In general,
the $W^\pm g$ fusion cross section decreases sharply as the scale increases,
whereas that of $e^+b$ events goes up mildly as $\mu$ gets larger (see 
Ref.~\cite{wuki} for a dedicated 
study in the case of $p\bar p$ collisions at the
Tevatron). Although at LO  there is no privileged choice
for $\mu$, Ref.~\cite{wgnloms} has shown that the most 
appropriate scale at the exact NLO (when both processes (\ref{CC}) and
(\ref{t}) need to be calculated) 
in the $b$ distribution function is $\mu^2\approx Q^2+m_t^2$, where
$Q^2\equiv-q^2$ 
($q$ being the four-momentum of the incoming virtual $W^\pm$ boson).
Therefore, we have adopted this value in producing 
Figure 2 (also as argument of the strong coupling constant), whereas
in all other cases we will maintain the LO `running' 
choice $\mu=\sqrt{\hat s}$. This has been done for two reasons.
First, we have
verified that for $\mu\OOrd \sqrt{\hat s}_{min} \approx m_t$ the rates
of process (\ref{t}) 
are rather stable, showing variations below 6-7\%. Second, this choice of
the scale allows one to consistently incorporate the non-resonant diagrams
along with the top ones when calculating the cross sections of the complete
processes (\ref{l})--(\ref{h_qcd}).  

From Figure 2, it is clear that,
apart from the different normalisation, the threshold behaviour in processes
(\ref{CC}) and (\ref{t})
is substantially similar as a function of the total CM energy. Though,
the ratio between the two series of curves is approximately 4.7-5.3 
at 300 GeV and  it
decreases  with increasing energy, stabilising at 1.3 TeV or so around
1.7--1.8. We trace back the behaviour at large energies as due to
the fact that the term $\alpha_s(\mu^2) \log(\mu^2/m_b^2)$ becomes 
constant because of large logarithms
cancelling each other ($\alpha_s(\mu^2)$ is in fact 
proportional to $1/\log(\mu^2/\Lambda^2_{\mbox{\tiny{QCD}}})$). In contrast,
at smaller energies (well below the TeV scale)
this is no longer the case and, in addition, graph 2
of Figure 1a becomes strongly suppressed, thus explaining the increase
of the observed ratio. The value of the latter between the
two cross sections when $\sqrt s_{ep}\OOrd1.3$ TeV
can be understood in terms of the large 
logarithms entering in the resummation of the $b$ structure function, which
tend to enhance the $b$ induced process with respect to the $g$ one.
For example, 
for $\mu^2=m_t^2$, with $m_t=170(175)[180]$ GeV, one gets the `leading logs'
$L=
\alpha_s(\mu^2)\log(\mu^2/m_b^2)\approx0.75(0.76)[0.76]$. Such  differences
between $b$ and $g\ar b\bar b$ induced processes at the TeV scale are 
not unusual in
 literature, see, e.g., Refs.~\cite{wuki,bvsg} (though,
for the case of hadron-hadron
 collisions at the TeV scale). Note that we obtain the same pattern
also for the case of on-shell top production, when no decay of the top
quark is implemented. 

Before proceeding, we should in fact mention that
we have studied the size of the differences
between the total rates of the two processes as obtained, on the one
hand, by using a finite width and implementing the decay currents and, on
the other hand, by keeping the top on-shell. In general, they are at the level
of few percent (the on-shell rates being larger). For example,
for the $b(g)$ induced process they vary between 2(1)\% to 4(5)\% when
$m_t=175$ GeV. In fact, rates are rather insensitive to the value of the
top mass. 

From Figure 2, we note in general
that although the cross section for process (\ref{t}) is small at existing
collider energies ($\sqrt{s}_{ep}\approx 300$ GeV at DESY leads to a total
cross section of less than 1 fb, which is negligible given the current
integrated luminosity of about 20 pb$^{-1}$ \cite{desy} at each of the two
experiments), it increases steeply near the TeV scale. At the
LEP2$\oplus$LHC scale it is easily observable at the `conservative'
luminosity of 100 pb$^{-1}$ \cite{LHC_exp}. 
There is however a sizeable dependence on the top quark mass, especially at low
energies: the cross
section being smaller for a phase space suppressed by a higher mass. (All
our results hereafter assume the central value of 175 GeV.)

Table I shows the cross section of the signal process (\ref{t}) evaluated
at the LEP2$\oplus$LHC energy, for
 twenty-four different sets of structure
functions.
The dependence is found to be approximately 20\%, 
with the maximum value of the total cross section differing from the
minimum value by 823 fb. 
We believe such theoretical uncertainty to be already at the present time
a reasonably small error so to motivate further and more detailed simulation 
studies (including hadronisation, detector effects, reducible background
\cite{LHC_top}) of single top phenomenology. To appreciate this we note
that the result obtained by adopting the old set EHLQ1 \cite{EHLQ}
(i.e., the one used in Ref.~\cite{LHC_top})
differs by that produced by MRS(A) in Table I by more than 50\% !

As a further check, we present
the Bjorken $x$ and the $Q$ dependence
of the cross section of process (\ref{t}) for a selection of PDFs, in
Figures 3a and 3b (respectively).
In particular, we have included results for some older, MRS(R1) and CTEQ(3M),
and some newer, MRRS(1) and CTEQ(4M), sets, as representative of the two 
approaches MRS and CTEQ, each of the pair being fitted to a similar set
of experimental data so to allow for a more consistent comparison.
Note that the normalisations of the curves are to 
unity, in order
 to enlighten the differential behaviours of these quantities, in addition to
their effects on the total rates (as was done in Table I). The clear message from
Figures 3a and 3b is that the differences between the two pairs of sets are 
very small (as can be appreciated in detail in the central inserts), 
typically a few percent over all the available kinematic range in $x$
and $Q$. Although we do not show the corresponding curves, we have verified
that such considerations also apply to the other PDFs considered here.
Thus, also at differential level the theoretical error on the rates of 
process (\ref{t}) due to the PDFs is well under control already at
present. 

Though it is beyond the scope of this study to trace back whether the 
differences in Table I (and
Figures 3a--b) among the various sets 
are due to the gluon structure function or to the $g\ar b\bar b$ splitting
 (which onsets the $b$ structure function),
it is worth mentioning that it could well be that by the time a future $ep$ 
collider will be running the uncertainties on the former will be so under 
control that one might attempt to distinguish between different dynamics
proposed for the latter. In this respect, it would be 
interesting to assess whether the differences between MRRS(1) (dashed line)
and CTEQ(4M) (dot-dashed line) in Table I and in 
Figures 3a and 3b are genuinely due
to the dedicated treatment of the threshold region $\mu\sim\mu_b$ performed
in Ref.~\cite{MRSCHM} or not \cite{nomass}. 
Clearly, this will require a tight control on all sources 
of experimental error, in particular of the actual value of the
$b$-tagging efficiency and of the hadronisation process of the quarks
at the TeV scale.

Figure 4 shows the differential distributions interesting for the final state
phenomenology. Those in combined jet
masses are sharply peaked at the top and $W^\pm$ masses well above the
irreducible noise, indicating that for
the hadronic case the jet masses can be used to clearly identify the top
decays. This feature is convenient both for the elimination of top events
from any other hadronic three-jet processes to which the former
may act as a background, and for the elucidation of the top quark
physics at $ep$ colliders in, for example, probing the $b$ quark
distribution function. 
The spectra in transverse momenta $p_T$ show that neither cuts in
$p_T$ nor cuts in $p_T^{miss}$ affect the total cross section
dramatically, whereas that of $\Delta R$, the azimuthal-pseudorapidity
separation defined by $\Delta R = \sqrt{(\Delta\phi)^2+(\Delta\eta)^2}$
(where $\phi$ is the azimuthal angle and $\eta$ the pseudorapidity)
indicates that the requirement of resolving the hadronic jets (or
the requirement of an isolated lepton in the leptonic case) severely
reduces the event rate.
The majority of events are found within $\Delta R\Ord 1.5$, which is about
90 degrees in the azimuthal angle. This is because the visible jets and
the lepton come from the energetic top quark. Thus, at lower energies the
azimuthal-pseudorapidity spread in the top quark decay products will be
larger and hence the requirement of such jet/lepton isolation not so stringent. 
The distribution of the missing transverse momentum in the leptonic case,
and more specifically the electronic case, is small at low missing $p_T$
and  indicates that only a small proportion of the events will
emulate neutral current events of the form $ep\rightarrow eX$. However,
those which do will form a potentially dangerous background to high $Q^2$
neutral current events, as the low $\Delta R$ mentioned above will
concentrate the electrons to the high $Q^2$ region. 

Table II and Figure 5 show the total cross section after the acceptance
cuts. The following LHC-like constraints were implemented (see \cite{LHC_top} for
alternative selection strategies): for the leptonic channel, $p_T^{\ell^+},
p_T^{b}>20$ GeV, $p_T^{miss}>10$ GeV and $\Delta R_{\ell^+,b}>0.7$; for
the hadronic channel, $p_T^{j,j'}, p_T^{b}>20$ GeV, $p_T^{miss}>10$ GeV
and $\Delta R_{j,j',b}>0.7$. We have not introduced any cuts on pseudorapidity, as the
particles were found to be all concentrated in the narrow
$|\eta|<2.5$ region even before
any selection in $p_T$ was made.

Table II summarises the event rates for all channels at three different
CM energies. The numbers in square brackets are the cross
sections of processes (\ref{b_barl}) and (\ref{b_barh_ew}). These are
additional backgrounds when bottom quark charge tagging is unavailable.
Since these effectively only differ from processes (\ref{l}) and
(\ref{h_ew}) in their non-resonant top quark production, they can be taken
as a measure of the magnitude of the irreducible background. As can be
noticed, 
such background effects are small. We see that the hadronic cross section
is higher at lower energy since, as discussed above, the acceptance cut in
$\Delta R$ affects the rates less at smaller values of $\sqrt s_{ep}$, thus
compensating for the reduced total cross section shown in Figure 2. 
From Figure 5 we see that background effects do not spoil the sharp resonances
in combined jet masses even after the acceptance cuts. We particularly 
stress that
the QCD background is negligible: luckily enough, as it curiously peaks around 
the $M_{W^\pm}$ value in the di-jet mass distribution. 
It can also be noted that the cut in $\Delta R$, 
the jet separation, of 0.7 is a conservative choice and, as can be
deduced from the distribution in $\Delta R$ in Figure 4, so that the rates 
should be expected 
much higher for looser constraints, especially for the hadronic channel when
three separate cuts in $\Delta R$ need to be made to resolve the three
jets completely.
Finally, given the optimistic vertex tagging performances foreseen for 
the LHC detectors \cite{higgs2}, we would expect that only a small fraction
of the event rates given in Table II will be lost in the actual analyses. 
In fact, we believe 
that the original capabilities of the LHC $\mu$-vertex devices 
will be maintained while running the CERN machine in the 
proposed $ep$ mode.

\section*{4. Summary and conclusions}

The single top quark production from initial state bottom sea quark at
future $ep$ colliders was studied, mainly focusing our attention to the
case of the proposed LEP2$\oplus$LHC accelerator
with the positron (electron) beam energy of 100
GeV and the proton one of 7 TeV.
The total cross section was found to be about 4 pb at this energy. 
The uncertainty  due to the structure functions was 
quantified to be rather small already at the present time, around 20\%, and 
is expected to diminish significantly before new $ep$ machines will enter into
operation. Furthermore, based on such a consideration,  
some optimistic prospects about the possibility of 
exploiting single top phenomenology in order to study the $g\ar b\bar b$ 
dynamics inside the proton were given.
Both the leptonic and hadronic decay channels of the top quark were
studied, in presence of the corresponding irreducible backgrounds, which have
been computed here for the first time. In the hadronic case,
distributions in the reconstructed top quark and $W^\pm$ boson masses were
found to be sharply peaked above the irreducible noise, so to allow for 
a prompt recognition of single top events. 
In both channels the cross section for the
background was found to be small compared with the signal events.  The
residual dependence of the latter
on the top quark mass was evaluated in several instances.
Finally, the formula for the matrix element squared of single top production, including 
top width effects and all the dynamic correlations between the top
decay products, was presented in order
to aid future, more detailed, experimental simulations. The complete numerical
programs, evaluating irreducible background effects as well, 
have been especially optimised
in view of high statistic Monte Carlo simulations and are available from the
authors upon request.

\section*{5. Acknowledgements}

SM is grateful to the UK PPARC and KO to Trinity College and the Committee
of Vice-Chancellors and Principals of the Universities of the United
Kingdom for financial support. SM also acknowledges the kind hospitality of
the Theoretical Physics Groups at Fermilab (USA) and Lund (Sweden), where
part of this work was carried out. Finally, the research performed in Lund 
by SM has been partially supported by the Italian Institute of 
Culture `C.M. Lerici' under the grant Prot. I/B1 690, 1997.

\goodbreak

\vfill
\newpage

\subsection*{Table Captions}
\begin{description}

\item{[I] } Total cross sections (hadronic and leptonic) for process
(\ref{t}) at LEP2$\oplus$LHC energies for twenty-four different sets of
structure functions.  Errors are
as given by VEGAS (the same statistics were used for the {\tt NCALL} and
{\tt ITMX} parameters) \cite{VEGAS}.

\item{[II] } Total cross sections (hadronic and leptonic, including
irreducible background effects) for processes (\ref{l})--(\ref{h_qcd})
at the LEP2$\oplus$LHC collider. The
structure function set MRS(A) was used. Errors are as given by VEGAS
\cite{VEGAS}. The following acceptance cuts were implemented:
(i) $p_T^{\ell^+}, p_T^{b}>20$ GeV, $p_T^{miss}>10$ GeV
and $\Delta R_{\ell^+,b}>0.7$ (leptonic channel);
(ii) $p_T^{j,j'}, p_T^{b}>20$ GeV, $p_T^{miss}>10$ GeV
and $\Delta R_{j,j',b}>0.7$ (hadronic channel).
In the squared brackets of the first two columns we report the rates of
the charge conjugates (\ref{b_barl})--(\ref{b_barh_ew}) of processes
(\ref{l})--(\ref{h_ew}), for which the resonant top production do not
occur. The rates of the charge conjugate of process (\ref{h_qcd}) are the
same as those in third column. 

\end{description}

\vfill
\newpage

\subsection*{Figure Captions}

\begin{description}

\item{[1] } Lowest order Feynman diagrams describing processes (\ref{CC}),
(\ref{l}), (\ref{h_ew}) and (\ref{h_qcd}), corresponding to sets 
({\bf a}), ({\bf b}), ({\bf c}) and ({\bf d}), respectively. Only the cases 
$\ell^+ \nu_\ell=e^+\nu_e$ and
$jj'=u\bar d$ are shown are shown for processes (\ref{l})--(\ref{h_qcd}),
whereas in reaction (\ref{CC}) the top is considered on-shell. The package 
MadGraph \cite{tim} was used to
produce the PostScript codes. In ({\bf c}) `A' represents the photon. The
dashed lines in ({\bf b}) and ({\bf c}) represent the SM Higgs boson and
the curly lines in ({\bf d}) the gluon. The number of diagrams in ({\bf
b}) reduces to 10 for the cases $\ell^+ \nu_\ell=\mu^+\nu_\mu$ and
$\tau^+\nu_\tau$, when diagrams 1, 2, 3, 5, 6, 7, 8, 12, 13, 20 and 21 do
not contribute.

\item{[2] } The total cross section (hadronic and leptonic channels) for
processes (\ref{t}) (upper lines) and
(\ref{CC}) (lower lines) for 300 GeV $\leq\sqrt{s}_{ep}\leq$ 2 TeV, 
with three different
values for the top quark mass: $m_t=170$ GeV (continuous line), $m_t=175$
GeV (dashed line) and $m_t=180$ GeV (dotted line). The structure function
set MRS(A) was used.

\item{[3] } Differential distributions in ({\bf a}) $x$ and 
({\bf b}) $Q$ for events of the type (\ref{t}) at the LEP2$\oplus$LHC 
collider for three representative sets of structure functions: 
MRS(R1) (solid), MRRS(1) (dashed), CTEQ(3M) (dotted) and CTEQ(4M) (dot-dashed).
In the central inserts, the spectra are magnified around their maximum
values. Normalisations are to unity.

\item{[4] } Differential distributions (hadronic and leptonic channels) 
for process (\ref{t}) at the LEP2$\oplus$LHC CM energy and
$m_t=175$ GeV in the following variables (clockwise). 
1.~$M_{\mbox{\tiny{jets}}}$,
the invariant mass of the two- (solid) and three-jet (dashed) systems in
hadronic decays. 2.~$p_T$, the transverse momenta of the lepton/jets
(solid) in leptonic/hadronic decays, of the bottom quark (dashed) in both
channels, and of the missing particles in leptonic (dotted) and hadronic
(dot-dashed) decays. 3.~$\Delta R$, the azimuthal-pseudorapidity
separation of the pairs lepton/jets-bottom quarks (solid) in
leptonic/hadronic decays.
The normalisation is to unity. The structure function set MRS(A) was used. 
In the case of the hadronic decays we have considered only one of the two
light quark jets, their distributions in the above variables being very
similar.

\item{[5] } Differential distributions in $M_{\mbox{\tiny{jets}}}$ (hadronic
channel only) for processes (\ref{h_ew}) (left) and (\ref{h_qcd}) (right)
at the LEP2$\oplus$LHC CM energy and $m_t=175$ GeV. 
$M_{\mbox{\tiny{jets}}}$ signifies the invariant mass of the two- (solid) and
three-jet (dashed) systems in hadronic decays. The normalisations are to
the total cross sections. The structure function set MRS(A) was used. Bins
are 2 GeV wide. The following acceptance cuts were implemented: 
$p_T^{j,j'}, p_T^{b}>20$ GeV, $p_T^{miss}>10$ GeV and $\Delta
R_{j,j',b}>0.7$ (hadronic channel). 

\end{description}

\vfill
\newpage

\begin{table}
\begin{center}
\begin{tabular}{|c|c|}
\hline
\multicolumn{2}{|c|}
{\rule[0cm]{0cm}{0cm}
single-top}
\\ \hline
\rule[0cm]{0cm}{0cm}
PDFs & $\sigma_{t}$ (fb) \\ \hline\hline
\rule[0cm]{0cm}{0cm}
MRS(A)   & $3817\pm15$ \\ 
MRS(A')  & $3736\pm14$ \\ 
MRS(G)   & $3607\pm14$ \\ 
MRS(J)   & $3860\pm15$ \\ 
MRS(J')  & $4159\pm16$ \\ 
MRS(R1)  & $3509\pm14$ \\ 
MRS(R2)  & $3815\pm14$ \\ 
MRS(R3)  & $3664\pm13$ \\ 
MRS(R4)  & $3868\pm15$ \\ 
MRS(105) & $3403\pm13$ \\ 
MRS(110) & $3606\pm13$ \\ 
MRS(115) & $3566\pm14$ \\ 
MRS(120) & $3844\pm15$ \\ 
MRS(125) & $3895\pm15$ \\ 
MRS(130) & $3967\pm15$ \\ 
MRRS(1)  & $4063\pm16$ \\ 
MRRS(2)  & $4070\pm16$ \\ 
MRRS(3)  & $4055\pm16$ \\ 
CTEQ(2M) & $3943\pm16$ \\ 
CTEQ(2MS)& $3840\pm15$ \\ 
CTEQ(2MF)& $3968\pm15$ \\ 
CTEQ(2ML)& $4226\pm17$ \\ 
CTEQ(3M) & $4193\pm16$ \\
CTEQ(4M) & $4108\pm15$ \\ \hline\hline
\multicolumn{2}{|c|}
{\rule[0cm]{0cm}{0cm}
no acceptance cuts}
\\ \hline

\hline\hline
\multicolumn{2}{|c|}
{\rule[0cm]{0cm}{0cm}
LEP2$\oplus$LHC}
\\ \hline

\multicolumn{2}{c}
{\rule{0cm}{1.0cm}
{\Large Table I}}  \\

\end{tabular}
\end{center}
\end{table}

\vfill
\newpage

\begin{table}
\begin{center}
\begin{tabular}{|c|c|c|c|}
\hline
\multicolumn{4}{|c|}
{\rule[0cm]{0cm}{0cm}
$\sigma_{tot}$ (fb)}
\\ \hline
\rule[0cm]{0cm}{0cm}
$\sqrt s_{ep}$ (TeV) & leptonic  & hadronic (EW)  & hadronic (QCD)  
\\ \hline\hline
\rule[0cm]{0cm}{0cm}
1.0   & $180.9\pm2.5[1.939\pm0.022]$ & $211.3\pm2.0[1.738\pm0.013]$ 
& $3.208\pm0.010$ \\ 
1.3   & $309.6\pm8.7[3.621\pm0.066]$ & $196.5\pm3.1[2.419\pm0.055]$ 
& $4.578\pm0.016$   \\ 
1.7   & $480.\pm12.[6.497\pm0.076]$  & $116.2\pm5.0[3.034\pm0.071]$ 
& $6.073\pm0.020$   \\ 
\hline\hline
\multicolumn{4}{|c|}
{\rule[0cm]{0cm}{0cm}
after acceptance cuts}
\\ \hline

\hline\hline
\multicolumn{4}{|c|}
{\rule[0cm]{0cm}{0cm}
MRS(A)}
\\ \hline

\multicolumn{4}{c}
{\rule{0cm}{1.0cm}
{\Large Table II}}  \\

\end{tabular}
\end{center}
\end{table}

\vfill
\clearpage

\begin{figure}[p]
~\epsfig{file=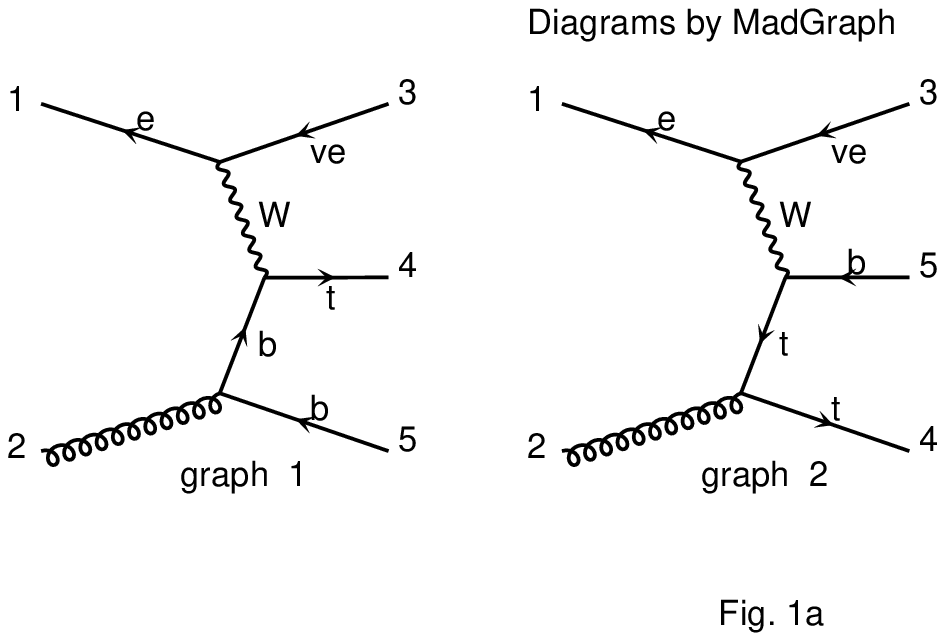,height=23cm}
\vspace*{2cm}
\end{figure}
\vfill
\clearpage

\begin{figure}[p]
~\epsfig{file=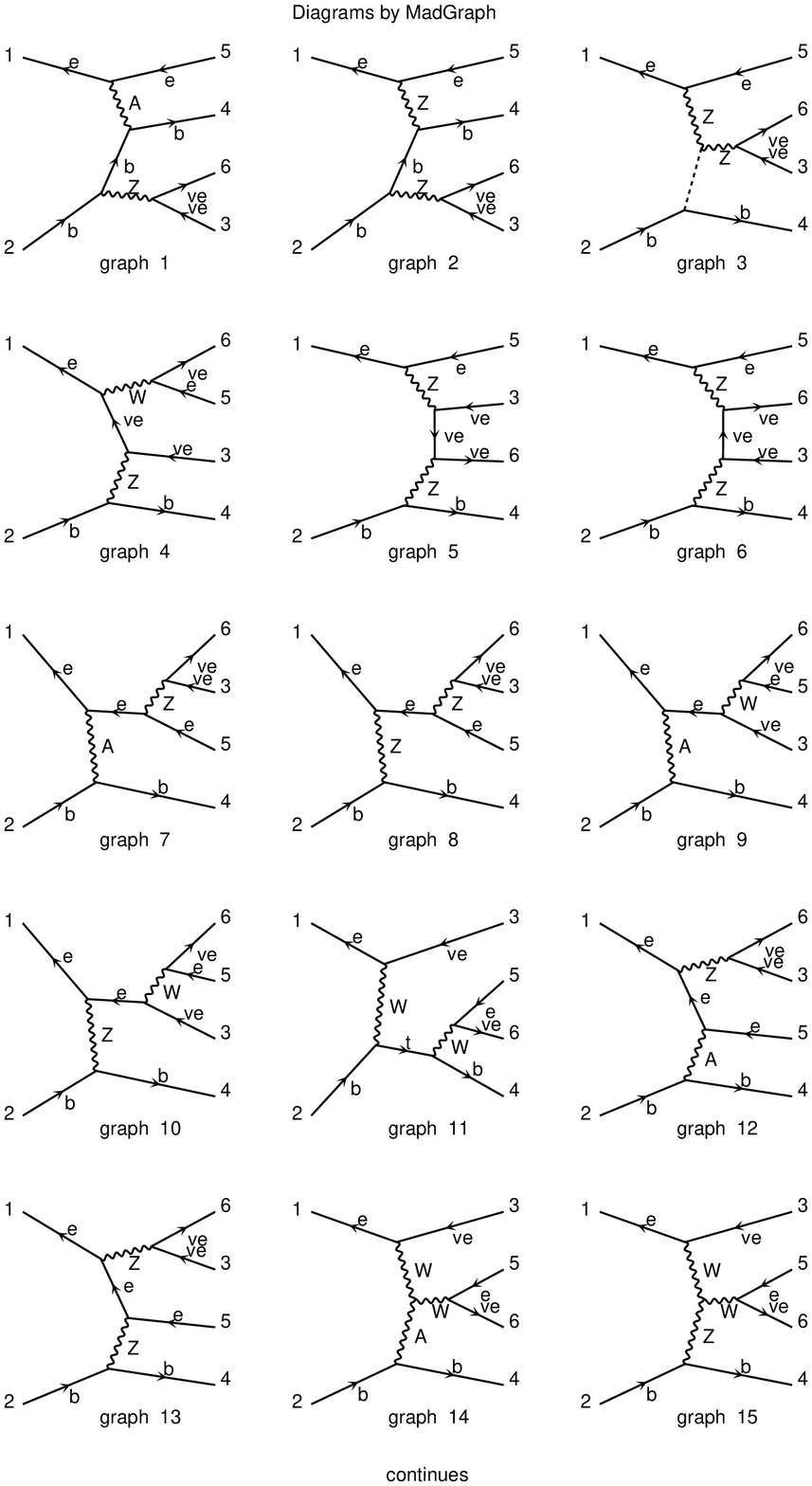,height=23cm}
\vspace*{2cm}
\end{figure}
\vfill
\clearpage

\begin{figure}[p]
~\epsfig{file=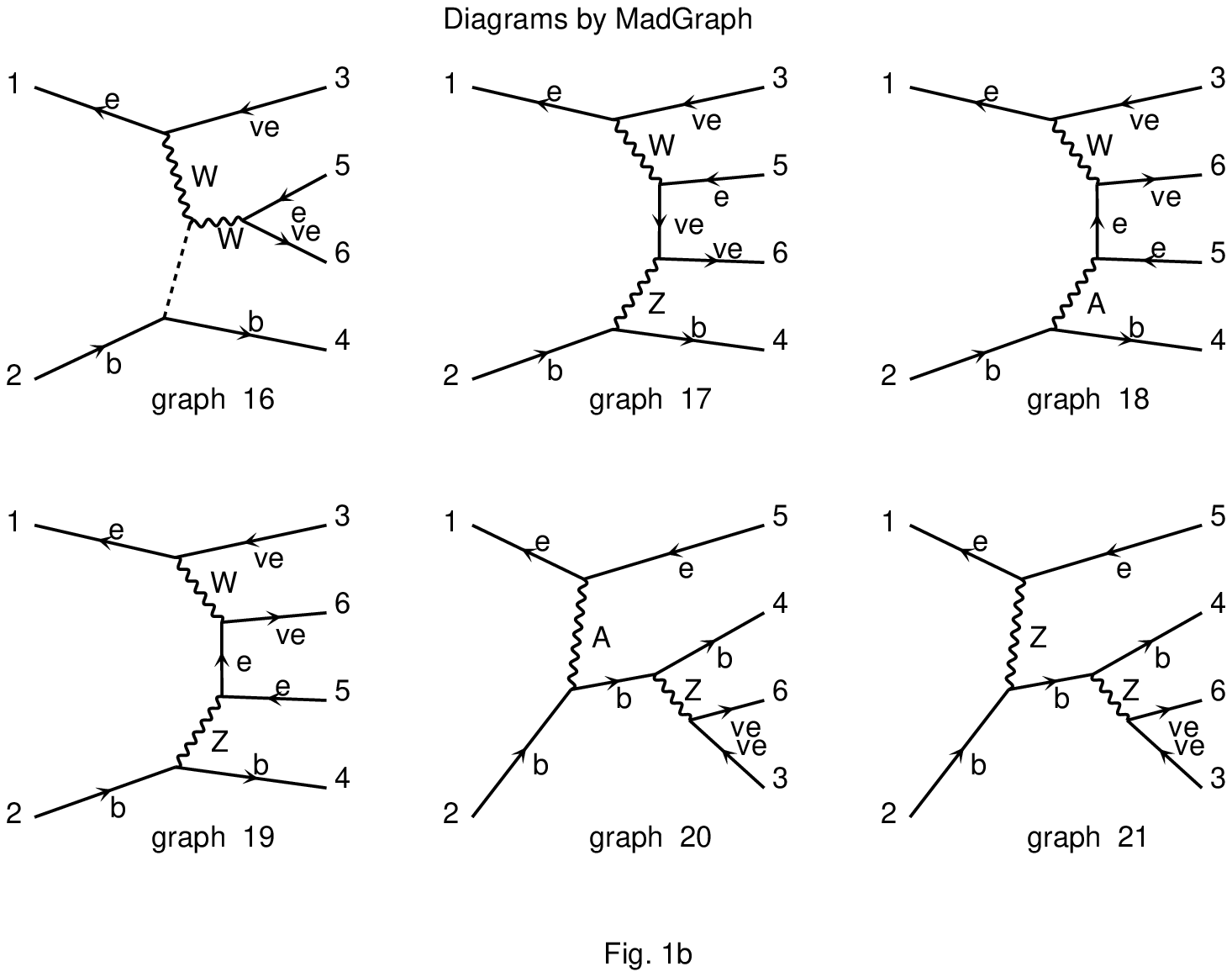,height=23cm}
\vspace*{2cm}
\end{figure}
\vfill
\clearpage

\begin{figure}[p]
~\epsfig{file=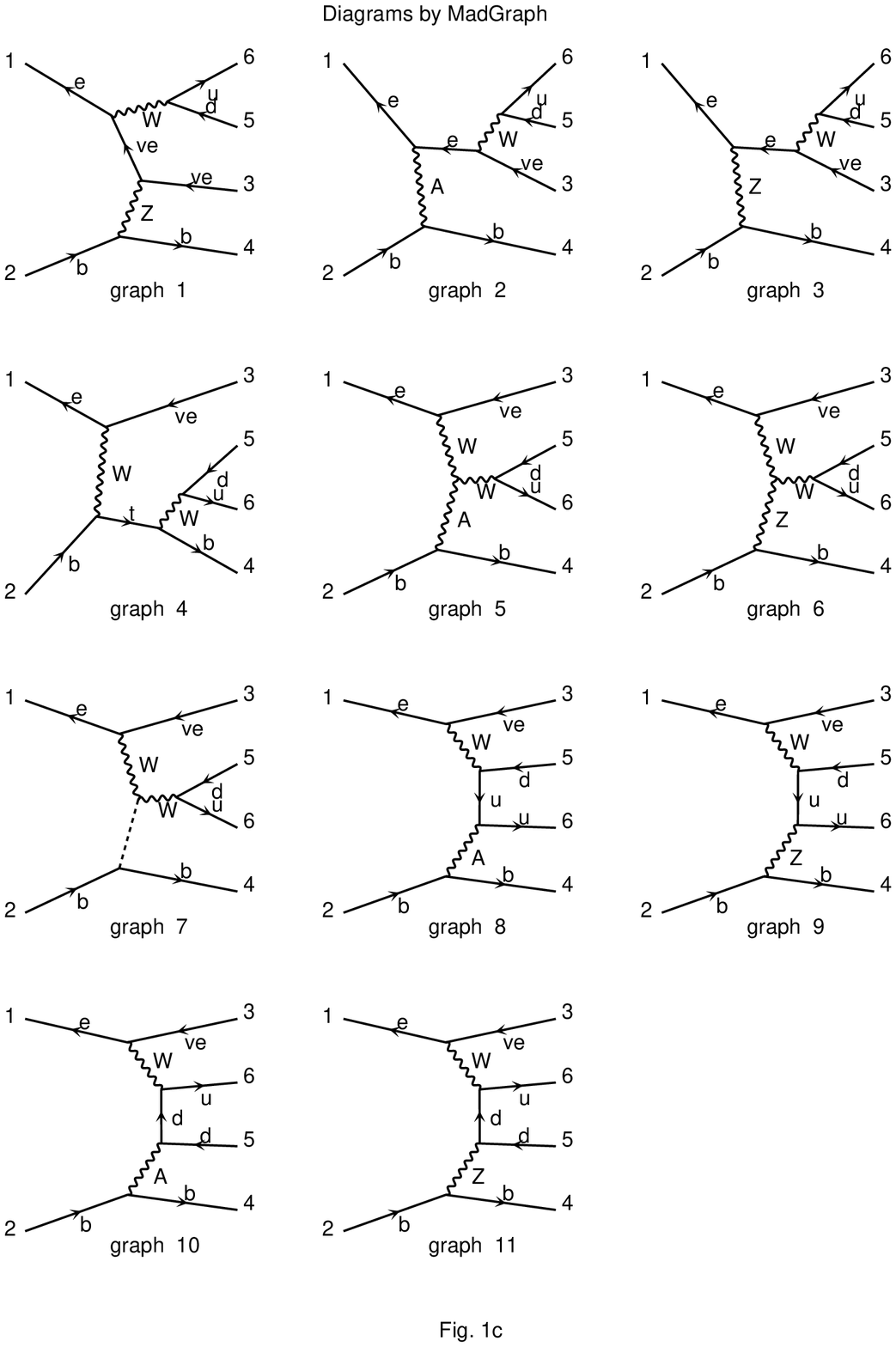,height=23cm}
\vspace*{2cm}
\end{figure}
\vfill
\clearpage

\begin{figure}[p]
~\epsfig{file=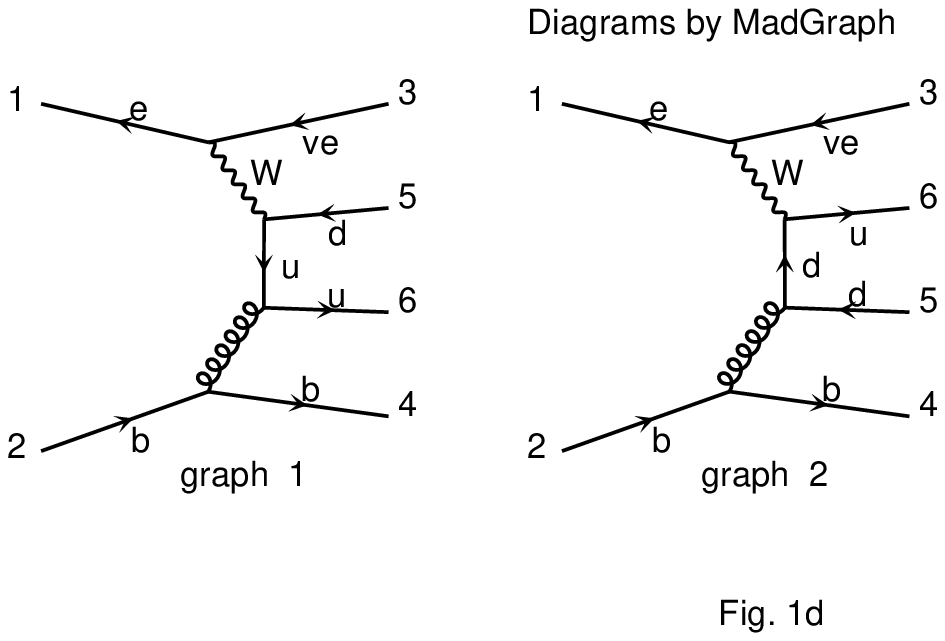,height=23cm}
\vspace*{2cm}
\end{figure}
\vfill
\clearpage

\begin{figure}[p]
~\epsfig{file=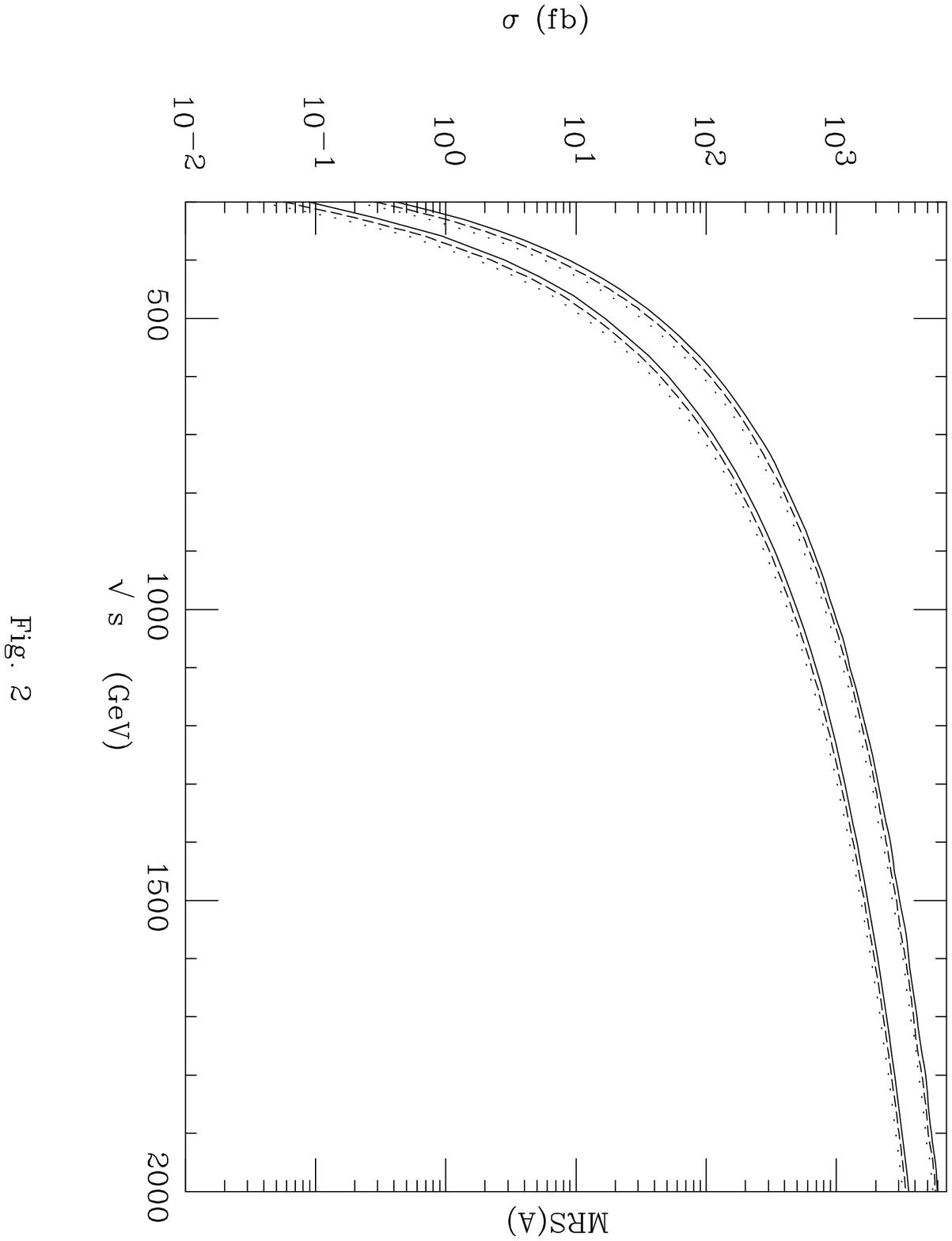,height=23cm}
\vspace*{2cm}
\end{figure}
\vfill
\clearpage

\begin{figure}[p]
~\epsfig{file=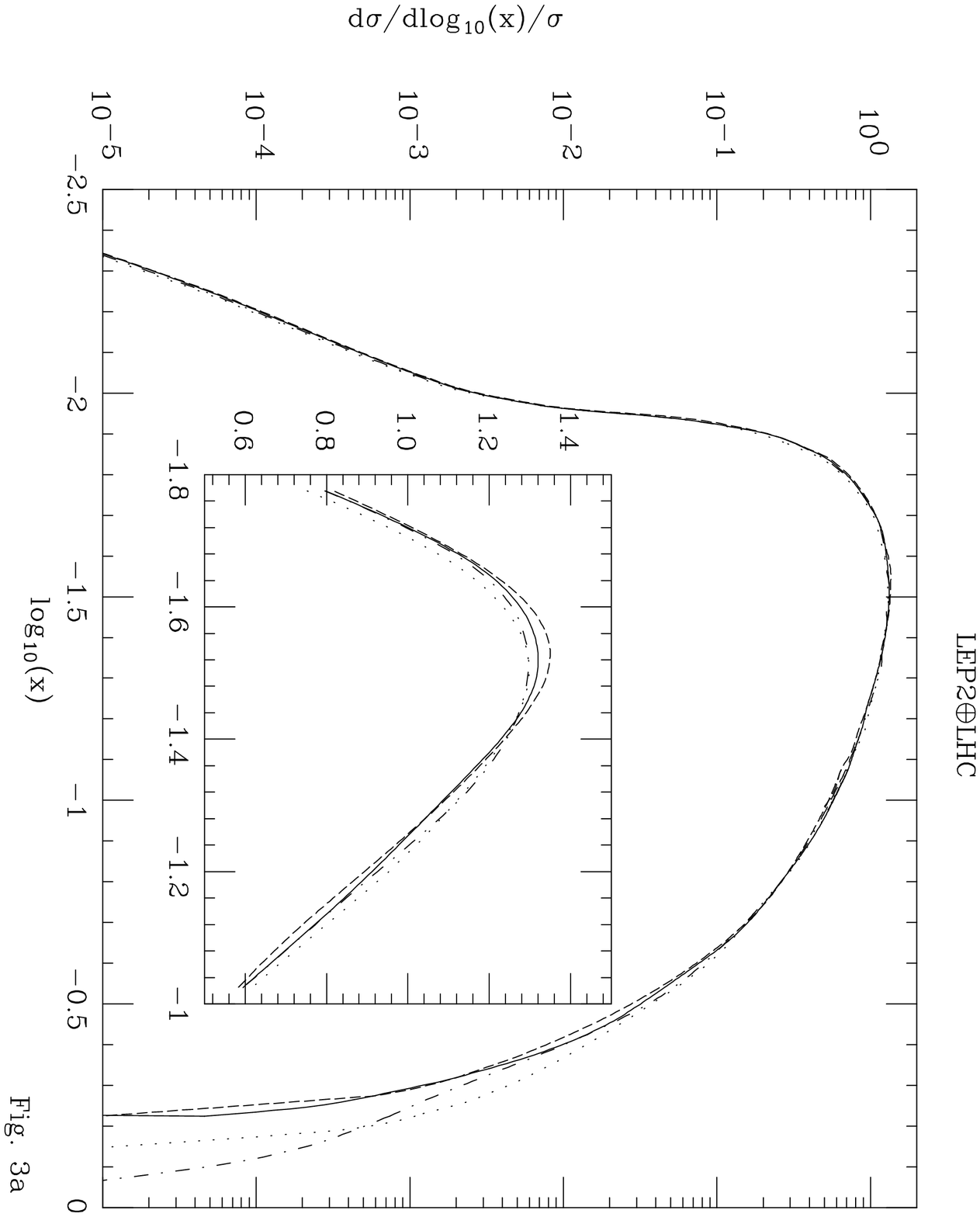,height=23cm}
\vspace*{2cm}
\end{figure}
\vfill
\clearpage

\begin{figure}[p]
~\epsfig{file=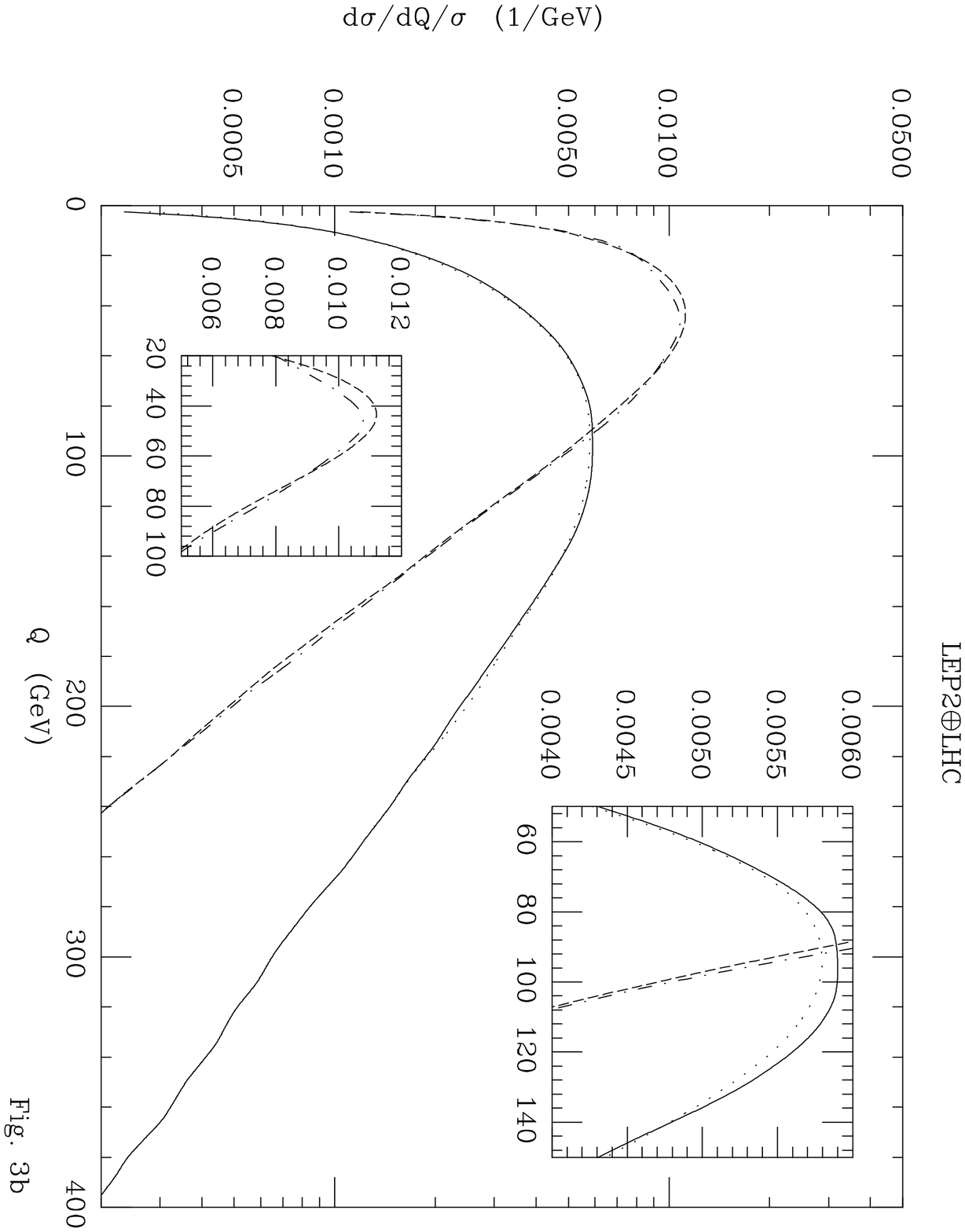,height=23cm}
\vspace*{2cm}
\end{figure}
\vfill
\clearpage

\begin{figure}[p]
~\epsfig{file=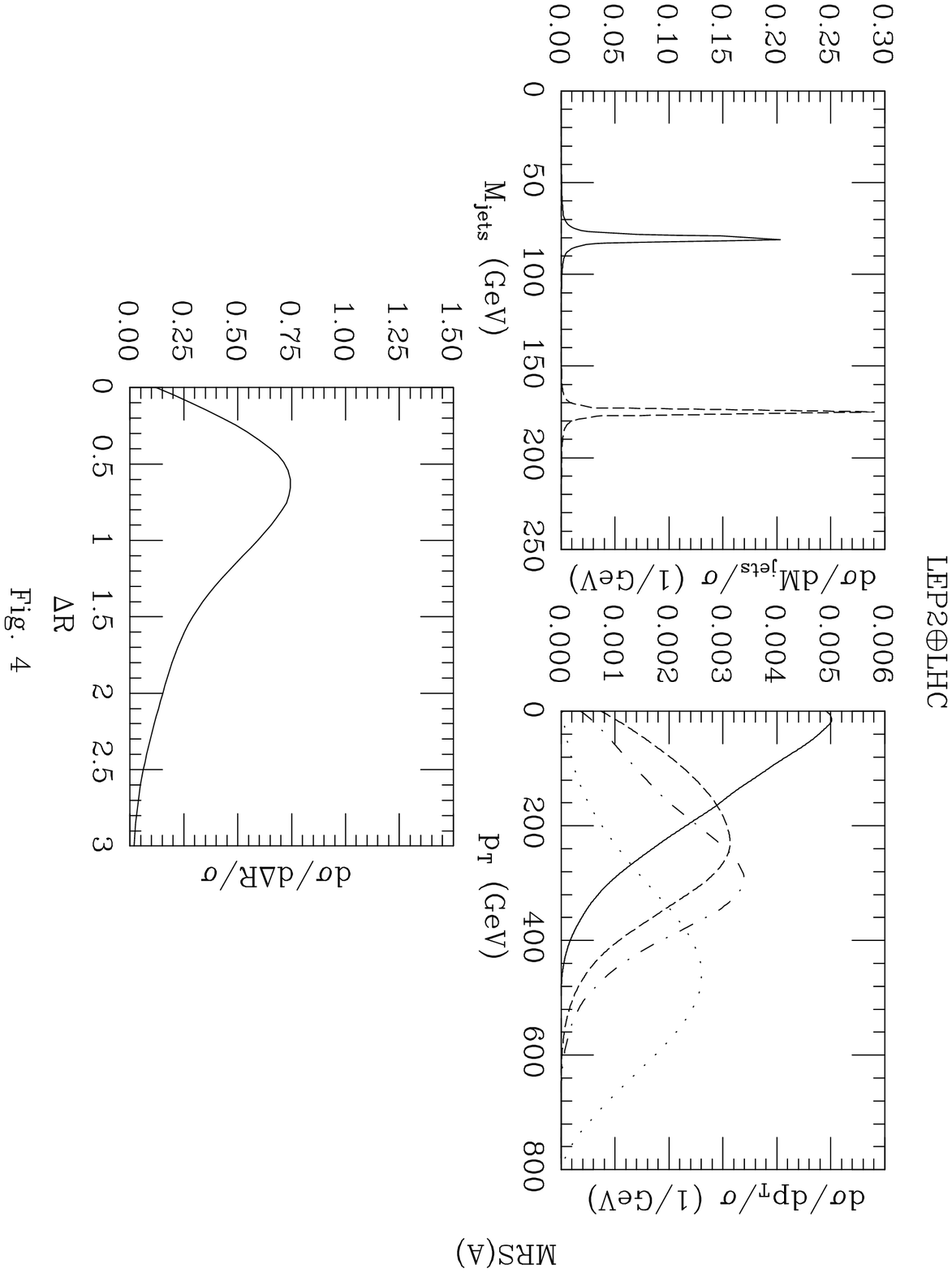,height=23cm}
\vspace*{2cm}
\end{figure}
\vfill
\clearpage

\begin{figure}[p]
~\epsfig{file=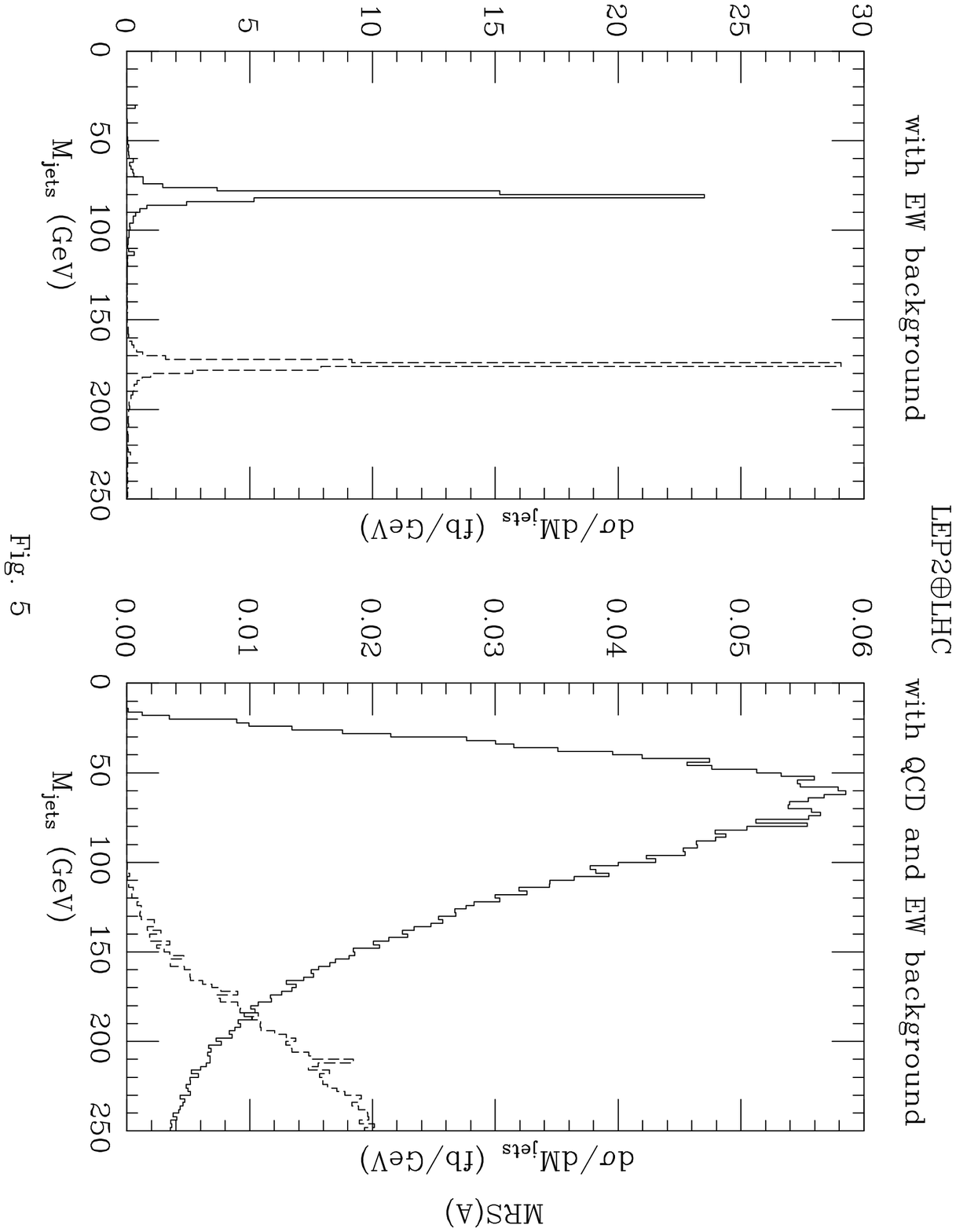,height=23cm}
\vspace*{2cm}
\end{figure}
\vfill
\clearpage

\vfill
\end{document}